\documentstyle[12pt,amssymbols]{article}
\def\ltap{\raisebox{-.4ex}{\rlap{$\sim$}} \raisebox{.4ex}{$<$}}

\def\journal{\topmargin .3in    \oddsidemargin .5in
        \headheight 0pt \headsep 0pt
        \textwidth 5.625in 
\textheight 8.25in 
        \marginparwidth 1.5in
        \parindent 2em
        \parskip .5ex plus .1ex         \jot = 1.5ex}
\journal

\newskip\humongous \humongous=0pt plus 1000pt minus 1000pt

\newif\ifdtup

\begin{document}
\begin{titlepage}
\begin{center}
March 27, 1995   \hfill    LBL- 36949\\
\hfill hep-ph/9503458
\vskip .5in

{\large \bf Probing for Ultraheavy Quanta at LHC }
\footnote
{This work was supported by the Director, Office of Energy
Research, Office of High Energy and Nuclear Physics, Division of High
Energy Physics of the U.S. Department of Energy under Contract
DE-AC03-76SF00098.}

\vskip .5in

Michael S. Chanowitz\footnote{Email: chanowitz@lbl.gov}

\vskip .2in

{\em Theoretical Physics Group\\
     Lawrence Berkeley Laboratory\\
     University of California\\
     Berkeley, California 94720}
\end{center}

\vskip .25in

\begin{abstract}
Experiments at the LHC are sensitive to the presence or absence of matter
quanta at mass scales far beyond the scales they can probe directly.
The production of $Z$ boson pairs by gluon-gluon fusion is greatly
enhanced if there are ultraheavy quanta that carry $SU(3)_{\rm Color}$ and
get their mass from electroweak symmetry breaking. For
example, a fourth generation quark doublet with an arbitrarily heavy mass
would induce a large excess in the $ZZ$ yield that could
be detected at the LHC with only $\simeq 10\%$ of the
design luminosity.
\end{abstract}

\end{titlepage}

\renewcommand{\thepage}{\roman{page}}
\setcounter{page}{2}
\mbox{ }

\vskip 1in

\begin{center}
{\bf Disclaimer}
\end{center}

\vskip .2in

\begin{scriptsize}
\begin{quotation}
This document was prepared as an account of work sponsored by the United
States Government. While this document is believed to contain correct
 information, neither the United States Government nor any agency
thereof, nor The Regents of the University of California, nor any of their
employees, makes any warranty, express or implied, or assumes any legal
liability or responsibility for the accuracy, completeness, or usefulness
of any information, apparatus, product, or process disclosed, or represents
that its use would not infringe privately owned rights.  Reference herein
to any specific commercial products process, or service by its trade name,
trademark, manufacturer, or otherwise, does not necessarily constitute or
imply its endorsement, recommendation, or favoring by the United States
Government or any agency thereof, or The Regents of the University of
California.  The views and opinions of authors expressed herein do not
necessarily state or reflect those of the United States Government or any
agency thereof, or The Regents of the University of California.
\end{quotation}
\end{scriptsize}

\vskip 2in

\begin{center}
\begin{small}
{\it Lawrence Berkeley Laboratory is an equal opportunity employer.}
\end{small}
\end{center}

\newpage

\renewcommand{\thepage}{\arabic{page}}
\setcounter{page}{1}

\noindent {\it \underline {Introduction} }

The matter sector is the least understood
part of the standard model.
No theoretical or experimental
constraint forbids the existence of additional quanta beyond the three
known quark-lepton families. The conventional wisdom that there are
no quarks heavier than the top quark is no more reliable
than the widely shared expectation of previous decades
that the top would not weigh more than 40 or 50
GeV. Provided they are too heavy to produce
at existing accelerators and that their weak $SU(2)_L$ multiplets are
sufficiently degenerate to satisfy the $\rho$ parameter
constraint,\cite{veltman,cfh} additional
ultraheavy quanta may be consistent with  precision
electroweak data. As many as two
ultraheavy quark-lepton families --- or other more exotic
varieties of matter quanta ---
are not excluded at the present level of precision.
The degree of mass degeneracy required
may seem unnatural, but final judgement would be premature
given our total ignorance of the origin of quark and lepton masses.\footnote{
For instance, the custodial $SU(2)$ of the symmetry breaking sector might
naturally apply to the fourth generation, with the lighter
fermions' masses generated by radiative corrections from
an extended gauge sector.}

The existence of ultraheavy quanta that carry $SU(3)_{\rm Color}$ and
obtain their mass from the
electroweak symmetry breaking condensate can be probed at the
LHC by means of their virtual loop contribution to the
process $gg \rightarrow ZZ$. This paper presents
the signals and backgrounds for the LHC at its 14 TeV design energy
and for the possible preliminary stage at 10 TeV. At 14 TeV the signal is quite
large: the increased $ZZ$ yield from one ultraheavy quark doublet
could be detected with only 10
fb$^{-1}$, little more than one month at the $10^{34}$cm$^{-2}$sec$^{-1}$
design luminosity. At 10 TeV and
$\simeq 3 \times 10^{33}$cm$^{-2}$sec$^{-1}$ observation would require
$1{1\over 2}$ to 2 years.

Is there an upper limit to the mass of matter quanta that get
their mass from the electroweak symmetry breaking condensate? We know
that the mass scale of the symmetry breaking sector is
constrained to be $\ltap O(2)$ TeV in order
to preserve the unitarity of $W$ and $Z$ boson
interactions (see for instance \cite{mc-mkg}).
No analogous limit constrains the mass of matter quanta.
The so-called ``unitarity upper
limit'' on quark and lepton masses\cite{cfh}, $\simeq 0.5$ and 1.0 TeV
respectively, is really just the mass scale at which tree
unitarity is saturated and higher orders become important, i.e., the onset of
strong Yukawa interactions. It does not mean that heavier quarks and leptons
are forbidden. An upper limit of order $\simeq 3$ TeV is
suggested,\cite{einhorn} based not on unitarity but on
dynamical considerations analogous to those advanced previously for the Higgs
boson mass.\cite{dashen-neuberger} I will consider
quark masses of 0.5 TeV and above. The signal is not very sensitive to the
mass:
for $m_Q \ge 1$ TeV it is already
within 10\% of the asymptotic $m_Q \rightarrow \infty$ value.

Ultraheavy quanta in degenerate $SU(2)_L$ multiplets would
not contribute to $\rho$ but would contribute to the parameter
$S$.\cite{peskin-takeuchi} For
instance an ultraheavy quark-lepton family would contribute $\sim +0.21$
to $S$ at one loop order in perturbation theory.
However this value is only reliable as an order of magnitude estimate
since higher order corrections from the strong
interaction of the ultraheavy quanta
with the Higgs sector are not perturbatively calculable.

The nominal experimental value for $S$ is negative but with large errors.
A recent analysis by Takeuchi\cite{takeuchi}
using $\alpha(M_Z)=1/129.1$ from \cite{swartz} yields
a less negative value than before, $S= -0.17 \pm 0.28$
(for $m_t=175$ and $m_H=300$ GeV),
that is consistent at the $\simeq 2\sigma$ level with up to
two ultraheavy families assuming 0.21 per family.
Because the central value is negative the constraint
is weaker than it seems. If true,
$S<0$ requires unknown nonstandard model
physics, since the standard model (and most other models) predicts $S>0$.
A negative contribution from new physics is {\it ab initio} of
unknown magnitude and could cancel a positive
contribution from ultraheavy quanta. On the other hand, if
$S$ is actually positive, the fit should include $S>0$ as a
constraint.\cite{peskin-takeuchi} Imposing $S>0$ and taking
$\alpha(M_Z)$ from \cite{swartz}, Takeuchi finds $S<0.44$ at 95\%
confidence,\cite{takeuchi}
again consistent with as many as two ultraheavy families using the one loop
value for $S$.

The analogous photon induced process, $\gamma \gamma
\rightarrow ZZ$, was considered previously, with the expectation
that the signal at a TeV photon
collider would be cleaner
than the gluon induced signal at a hadron
collider.\cite{mc_gamma} But the $W$ boson loop amplitude
was later found to contribute a large
background\cite{W-loop} that buries
the signal for $\sqrt{s_{\gamma \gamma}} \leq 3$ TeV.
The absence of the $W$ loop background
is a great advantage for the
$gg \rightarrow ZZ$ process.

The nondecoupling of ultraheavy quanta in $gg \rightarrow ZZ$ was noted
by Glover and van der Bij\cite{glover-vanderbij}. It was considered
by Hagiwara and Murayama\cite{hm}
(a fact not known to the author when \cite{mc_gamma} was written), using
a different method, in R-gauge rather than U-gauge.
Their paper had a different emphasis, focusing
on multiple weak boson production at the SSC in the asymptotic
$m_Q \rightarrow \infty$ limit. They did not consider the backgrounds (or the
interference of signal and background
amplitudes) nor did they consider LHC collider energies.

The purpose of the present paper is
to establish how well the $ZZ$ signal can be seen at the
LHC, taking account of the
backgrounds from $\overline qq$ annihilation, $gg$ fusion, and the order
$\alpha_W^2$ amplitude $qq \rightarrow qqZZ$.  Experimental cuts
are presented that optimize the signal relative to the
background. Signal cross sections are considered for $m_Q$
between 0.5 and 10.0 TeV and for the $m_Q \rightarrow \infty$ limit,
including the coherent interference of the signal and background
$gg \rightarrow ZZ$ amplitudes.

Since the $HXX$ coupling is strong for ultraheavy quanta $X$
that obtain their mass from the Higgs boson, higher order Higgs
boson exchange corrections  are not under perturbative control.
Consequently the one loop signal amplitudes can
only indicate the order of magnitude, and the cross sections
do not precisely probe the quantum numbers or the number of
ultraheavy quanta. The same
limitation applies to $\gamma \gamma \rightarrow ZZ$, to the on-shell
$H \rightarrow \gamma \gamma$ and $Z \rightarrow H \gamma$ partial widths,
and to the electroweak parameter $S$.

The next sections review the basic physics, present signal and
background events rates  for the optimal experimental cuts,
and discuss the results.

\noindent {\it \underline {One Loop Amplitude for Ultraheavy Quanta }}

There are two important features: 1) that ultraheavy
quanta $X$ do not decouple and 2) that $\sigma(gg \rightarrow ZZ)$
increases linearly with $s$ in the domain
$$
m_X^2 \ \gg \ s \ \gg \ m^2_H. \eqno(1)
$$
They are seen most easily in unitary gauge, for which the dominant contribution
is the triangle amplitude in figure 1.
The $X$ contribution does not decouple as
$m_X \rightarrow \infty$ because factor(s)
 of $m_X$ from the $HXX$ vertex cancel
factor(s) of $1/m_X$ from the loop integral.
\footnote{There are one or two factors
of $m_X$ for spin 1/2 and 0 respectively.
The leading $ggH$ off-shell amplitude is determined by the leading order trace
anomaly for a theory with an $SU(3)$ symmetry\cite{su3traceanomaly}, and the
QCD
corrections are precisely the higher order terms in the beta
function\cite{highordertrace}. However the QCD corrections are much smaller
than
the unknown higher order corrections from the Higgs sector.}
The energy dependence is understood as follows:
a factor $s$ from the $G^{\mu\nu}G_{\mu\nu}H$ structure of the $ggH$ vertex
(required by gauge
invariance), a factor $\simeq 1/s$ from the Higgs boson propagator, and a
factor $s$ from the U-gauge $HZZ$ vertex for longitudinally
polarized $Z$ bosons.

The leading amplitude mediated by ultraheavy quanta $X$ is then
$$
{\cal M} ( g^a_1g^b_2\rightarrow Z_LZ_L)_X =
  {S_X C_X  \alpha_S(s) \over 3\pi}
\ {s\over v^2}\ \delta_{ab} \delta_{\lambda_1 \lambda_2}\eqno(2)
$$
where $v=246$ GeV, $\alpha_S$ is the strong interaction coupling constant,
$a,b$ are color indices, $\lambda_i$ denote gluon polarizations,
and the subscript $L$
denotes longitudinal polarization. The spin factor is
$S_X = 1$ for spin 1/2 and $=1/4$ for spin 0. The $SU(3)$ quadratic Casimir
operator $C_X$ is normalized to 1/2 for $X$ in the triplet,
$C_X\delta_{ab}={\rm Tr}(T^a_X T^b_X)$. In U-gauge the box graph
amplitudes, figure 1, are suppressed relative to equation 2 by $s/m_X^2$.

Assuming $n_D$ ultraheavy quark doublets, the color- and spin-averaged
differential cross section following from equation 2 is
$$
{ d\sigma \over d {\rm cos}\theta} =
{\beta n_D^2 \over \pi}\
\left({\alpha_S  \over 96 \pi} \right)^2 {s\over v^4}\eqno(3)
$$
where $\theta$ is the polar scattering angle and $\beta$ is the $Z$ boson
velocity in the $ZZ$ center of mass. The signals presented below
for finite $m_Q$ and for $m_Q \rightarrow \infty$
also include the (constructive) interference of the
$X$-mediated loop amplitude and the background $gg \rightarrow ZZ$
amplitudes mediated by the three known quark doublets.\footnote{The
interference is
mostly from the $t$ quark amplitudes, though lighter quark loops make
significant contributions to the background.}

Using the R-gauge and the equivalence theorem it is easy to see
that equation 2 cannot follow from the triangle amplitude, figure 1,
since the $Hzz$ vertex $\sim m_H^2/v$ is negligible relative
to the $HZ_LZ_L$ vertex $\sim s/v$. The explanation is that the box graphs
provide the leading result in R-gauge. This has been verified by explicit
computation using an effective Lagrangian\cite{hm} and
by a general argument sketched in \cite{mc_gamma}.

\noindent {\it \underline {Cross Sections and Cuts}}

To maximize the yield we consider the ``silver-plated'' channel, first
suggested for heavy Higgs boson detection,\cite{mc-mkg,rnc-mc}
$ZZ \rightarrow \overline ll + \overline \nu \nu$,
where $l$ denotes an electron or muon.
The signature is a high $p_T$ $Z$
boson balanced by missing transverse energy.
This channel provides six times
more events than the ``gold-plated'' channel, $ZZ \rightarrow \overline ll
+ \overline ll$, and is nearly as clean for large transverse momentum.
It seems viable at the LHC according to both
ATLAS\cite{atlas} and CMS\cite{cms}. Even with 40 events per crossing,
the pile-up background is negligible for $E_T^{\rm miss} >
100$ GeV (see figure 11.15 of \cite{atlas}). The optimal cuts presented below
require transverse momenta more than twice as big, typically $\geq 250$ GeV.

The cross sections for the four charged lepton
channel can be estimated by dividing the cross
sections presented below by $\simeq 6$. Even though it
contains more information, it is not possible
to improve the signal:background ratio dramatically beyond what is achievable
for the two charged lepton final state. The results presented below are
conservative in that the $\simeq 15\%$ contribution
of the four charged lepton channel is not included.

The leading background is $\overline qq \rightarrow ZZ$. The second
background is $gg \rightarrow ZZ$ mediated by loop amplitudes of the six
known quarks (figure 1).\cite{glover-vanderbij}
For the optimal cuts the $gg \rightarrow ZZ$ background
is $\sim 15\%$ of the total background.
Another potential background is
the order $\alpha_W^2$ amplitude $qq \rightarrow
qqZZ$,\cite{qq-qqzz} computed in the standard
model assuming a light Higgs boson, $m_H \leq 100$ GeV. It includes $WW$ and
$ZZ$ fusion graphs as well as diagrams in which one or both
$Z$'s are radiated from an external quark line. It is potentially
larger than the $gg \rightarrow ZZ$ background but
a central jet veto (CJV) reduces it by an order of magnitude to a negligible
level, of order 1\% of the total background.

The CJV also suppresses the large NLO
(next-to-leading-order) background from
$ qg \rightarrow qZZ$.\cite{ohnemusetal} With the CJV
the lowest order $\sigma(\overline qq \rightarrow ZZ)$ cross section is
slightly
larger than the NLO inclusive $ZZ$ cross section, so that our use of
the lowest order $\sigma(\overline qq \rightarrow ZZ)$ is actually
a conservative background estimate.

The signal is distinguished from the background by three characteristics.
\begin{itemize}
\item The subprocess cross section increases with energy for the signal and
falls for the backgrounds.
\item The dominant background is peaked in the forward direction while the
signal is relatively isotropic.
\item The signal consists of
longitudinally polarized $Z$ boson pairs while the
background is dominated by $Z$ pairs with one or both $Z$'s
transversely polarized.
\end{itemize}

These features dictate the cuts. The first implies a cut on the
transverse momentum of the observed $Z$. The second suggests a central rapidity
cut, which is in any case required by the geometry of the detectors. The
first and third can be simultaneously exploited by
a cut on the transverse momentum of the $Z$ decay products, as noted
in studies of $W^+W^+$ scattering.\cite{wpwp} Since
longitudinally polarized $Z$'s tend to decay at right angles to the $Z$ line of
flight, both leptons typically share the transverse momentum of the parent
boson. For transversely polarized $Z$'s the decay tends to be along the
$Z$ line of flight, so that there is an unequal division of the parent $p_T$
and a greater likelihood that the softer lepton will fail a $p_{Tl}$ cut.

We define a conservative criterion for observability:
$$
\sigma^{\uparrow}   =  S/\sqrt{B}  \ge  5/\sqrt\epsilon \eqno(4)
$$
$$
\sigma^{\downarrow}   =  S/\sqrt{S+B}  \ge  3/\sqrt\epsilon \eqno(5)
$$
$$
S \ge B, \eqno(6)
$$
where $S$ and $B$ are the number of
signal and background events assuming 100\% detection
efficiency, and $\epsilon$ is the
experimental efficiency, assumed below to be 95\% for an isolated, high $p_T$
$Z$ decaying to $e^+e^-$ or $\mu^+\mu^-$.\cite{sdcloi} The requirement $S\geq
B$
is conservatively imposed to allow for theoretical uncertainty in the
magnitude of the background, probably $\leq 20$--$30\%$ after ``calibration
measurements'' of standard processes at the LHC.

The cuts are optimized over a
three dimensional parameter space consisting of
$p_{TZ}^{\rm MIN}$, $p_{Tl}^{\rm MIN}$, and $\eta_l^{\rm MAX}$.
The optimum cut is the one that satisfies equations 4--6 with the smallest
integrated luminosity, denoted ${\cal L}_{MIN}$.
In addition a central jet veto is imposed to reject
events containing one or more jets with $\eta_J < 3$ and $p_{TJ}>50$ GeV.

We consider the signal from one ultraheavy quark doublet of mass $m_Q$.
The integrated $p_{TZ}$ distribution for 14 TeV and 100 fb$^{-1}$ is shown in
figure 2, where $p_{Tl}>90$ GeV and
$\eta_l < 2$ are imposed. The background is indicated by the dashed line
while the coherent sum of signal
and background is shown in the solid lines for (from bottom to top)
$m_Q= 0.5$, 1.0, 2.0 TeV and
$m_Q \rightarrow \infty$. The asymptotic cross section is approached rapidly
from below: the $m_Q=1$ TeV signal is already within 10\% of the
$m_Q \rightarrow \infty$ limit.

The optimized signals and cuts are shown in tables 1 and 2 for
$\sqrt{s}=14$ and 10 TeV respectively, with $m_Q = 0.5$, 1.0, 2.0, 4.0,
and 10.0 TeV as well as $m_Q \rightarrow \infty$, the latter combined both
coherently ($\infty_C$) and incoherently ($\infty_I$) with the background.
While values of $\eta_l^{\rm MAX}$ from 1.0 to 2.5 were explored,
the rapidity cut is fixed at
$\eta_l^{\rm MAX}=2$ for the quoted results, because ${\cal L}_{MIN}$
is not very sensitive to variations between 1.75 and 2.5 and because the
detectors are likely to be most efficient for $\eta_l < 2$,

We see from table 1 that a signal satisfying equations 4--6 can be obtained
with $\simeq 10$ fb$^{-1}$, only 10\% of a year
at the design luminosity. For the optimal cuts the signal
is typically twice as large as the background.
The incoherent approximation, denoted by $\infty_I$, underestimates the
true signal by $\sim 20\%$.

For $\sqrt{s}=10$ TeV a significant signal requires 50--60 fb$^{-1}$ or
$1{1\over 2}$--2 years of running at the projected
3--4$\times 10^{33}$cm$^{-2}$sec$^{-1}$ luminosity. The signal:background
ratio for the optimal cuts falls to $\simeq$ 1:1.
Nevertheless the signal is big enough that it might be observable at 10 TeV.

\noindent {\it \underline {Discussion}}

In the analysis presented here we assumed a light Higgs boson and used
$m_H=100$ GeV in the computations. The results do not depend on the precise
value of $m_H$ as long as it is not heavier than a few hundred GeV. The
situation could be more complicated if $SU(2)_L$ breaking were due to
more than one Higgs boson or to a
strongly coupled Higgs boson (say $m_H \simeq 1$ TeV) or if it
were due to dynamical symmetry breaking. These complications would effect
the details but in each case a large signal would be expected, unless
different scalars generated the gauge boson and
ultraheavy masses.

Compared to other TeV scale gauge boson pair signals, the signals presented
here
are large. For the same cuts (including the CJV) the asymptotic signal in table
1 is 70 \% bigger than the $ZZ$ signal from the 1 TeV Higgs boson and three
times bigger than the strong scattering signal of the linear
model.\cite{mbmczz}

If an excess $ZZ$ signal were observed in longitudinally polarized pairs, the
interpretation would not be immediately clear. The magnitude of the signal
might
be a clue, especially if independent evidence for a light Higgs sector were
available. Strong $WW$ scattering would give rise to excesses also in the $WZ$
and/or $W^+W^+$ channels, while the $gg$ fusion signal of ultraheavy quanta
only contributes to $ZZ$ and $W^+W^-$. Jet tagging would also help to
distinguish since Higgs sector physics would
contribute to both\footnote{Note that
electroweak symmetry breaking by strong interactions above 1 TeV also
enhances $gg \rightarrow ZZ$ --- see \cite{mbmczz}.}
$gg \rightarrow ZZ$ and $qq \rightarrow qqZZ$ while
virtual ultraheavy quanta only enhance the former.
Measurements of $H \rightarrow \gamma \gamma$, $Z \rightarrow H \gamma$, and
the electroweak parameter
$S$ could provide corroborating evidence
but suffer from incalculable higher order corrections discussed above.
To confirm the interpretation of a signal we would eventually have to
observe the ultraheavy quanta directly.

A negative result would be easier to interpret. If no $ZZ$ excess were seen
beyond what could be accounted for by the Higgs sector,
we could conclude that ultraheavy quanta with
masses generated by electroweak symmetry breaking probably do not exist even at
arbitrarily high mass scales, also very useful information. We conclude that
experiments at the LHC are sensitive to the presence or absence of matter
quanta at mass scales far beyond the scales they can probe directly.

\vskip .2in
\noindent Acknowledgements: I wish to thank T. Takeuchi for helpful discussions
and for providing the results of his updated fits.
This work was supported
by the Director, Office of Energy
Research, Office of High Energy and Nuclear Physics, Division of High
Energy Physics of the U.S. Department of Energy under Contracts
DE-AC03-76SF00098 and DE-AC02-76CHO3000.

\newpage
\noindent {\bf Table 1} \\
\noindent Optimized yields for one ultraheavy quark doublet of mass $m_Q$,
for $\sqrt{s} = 14 $ TeV and $\eta^{\rm MAX}_\ell =2$.
For each $m_Q$, ${\cal{L}}_{\rm MIN}$ is the smallest integrated
luminosity that satisfies
equations 4-6, $S/B$ are the resulting numbers of
signal/background events per 100 fb$^{-1}$, and
$ p_{Tl}^{\rm MIN}, p_{TZ}^{\rm MIN}$
indicates the corresponding optimal cut.
$\infty_C$ and $\infty_I$ denote the $m_Q \to \infty$ limit combined
coherently or incoherently with the background.

\begin{center}
\vskip 10pt
\begin{tabular}{cccc}
$m_Q$(TeV) & ${\cal{L}}_{\rm MIN}$(fb$^{-1})$ &$S/B$(100 fb$^{-1})$
& $p_{Tl}^{\rm MIN}, p_{TZ}^{\rm MIN}$(GeV)\cr
\hline
&&&\cr
0.5 &17.4 &130/111 &70,250\cr
1.0 &12.2&121/67 &70,300  \cr
2.0&10.8&136/74&60,300\cr
4.0& 10.1  &149/85 &90,250\cr
10.0&9.9&150/85&90,250\cr
$\infty_C$&9.9&150/85&90,250\cr
$\infty_I$& 12.7&118/67&70,300

\end{tabular}
\end{center}

\vskip .5in
\noindent {\bf Table 2} \\
\noindent Results for $\sqrt{s} = 10 $ TeV and $\eta_l^{\rm MAX} =2$,
tabulated as in table 1 except that $S/B$ denotes the numbers of
signal/background events per 30 fb$^{-1}$.

\begin{center}
\vskip 10pt
\begin{tabular}{cccc}
$m_Q$(TeV) & ${\cal{L}}_{\rm MIN}$(fb$^{-1})$ &$S/B$(30 fb$^{-1})$
& $p_{Tl}^{\rm MIN}, p_{TZ}^{\rm MIN}$(GeV)\cr
\hline
&&&\cr

 0.5 &82 &9.4/9.2 &80,300\cr
1.0 &62&12.1/11.6 &100,250  \cr
2.0&52&14.3/13.6&90,250\cr
4.0& 49 &16.0/15.8 &80,250\cr
10.0&48&16.1/15.8&80,250\cr
$\infty_C$&48&16.1/15.8&80,250\cr
$\infty_I$& 66&7.7/5.0&90,350

\end{tabular}
\end{center}

\newpage
\begin{center}
{\bf Figure Captions}
\end{center}
\vskip .5 in
\noindent {\bf Figure 1} \\
\noindent Triangle and box diagrams for $gg \rightarrow ZZ$.

\vskip .5 in
\noindent {\bf Figure 2} \\
\noindent Numbers of events with
$p_{TZ}>p_{TZ}^{\rm MIN}$ for $\sqrt{s}=14$ TeV and 100 fb$^{-1}$.
Additional cuts are $\eta_l < 2$ and $p_{Tl}>90$ GeV. Signals are for
one ultraheavy quark doublet of mass $m_Q$.
The dashed line is the background,
and the four solid lines are, from bottom to
top, for $m_Q=0.5,\ 1.0,\ 2.0$ TeV and $m_Q \rightarrow \infty$.




\begin{thebibliography}{99}
\bibitem{veltman} M. Veltman, Nucl. Phys. {\bf B123:89},1977.
\bibitem{cfh} M. Chanowitz, M. Furman,
I. Hinchliffe, Phys. Lett. {\bf 78B:285},1978; Nucl. Phys. {\bf B153:402},
1979.
\bibitem{mc-mkg} M. Chanowitz and M. Gaillard, Nucl. Phys. {\bf B261:379},1985.
\bibitem{einhorn} M. Einhorn, Phys. Rev. Lett. {\bf 57:2115},1986.
\bibitem{dashen-neuberger} R. Dashen and H. Neuberger, Phys.Rev.Lett. {\bf
50:1897},1983.
\bibitem{peskin-takeuchi} M. Peskin and T. Takeuchi,
Phys.Rev.Lett. {\bf 65:964},1990 and Phys.Rev.
{\bf D46:381},1992.
\bibitem{takeuchi} T. Takeuchi, private communication.
\bibitem{swartz}  M. Swartz, SLAC-PUB-6710 (hep-ph 9411353), 1994.
\bibitem{mc_gamma} M. Chanowitz, Phys. Rev. Lett. {\bf69:2037},1992.
\bibitem{W-loop} G.V. Jikia, Phys. Lett. {\bf B298:224},1992 and
Nucl. Phys. {\bf B405:24},1993;
M.S. Berger, Phys. Rev. {\bf D48:5121},1993;
B. Bajc, Phys. Rev. {\bf D48:1903}, 1993; D.A. Dicus
and C. Kao, Phys. Rev. {\bf D49:1265},1994; H. Veltman, Z.Phys. {\bf C62:235},
1994.
\bibitem {glover-vanderbij} E. Glover, J. van der Bij, Nucl. Phys.
{\bf B321:561},1989. The
evaluation of the amplitude for $s \ll m_X^2$, equation (5.4), is too large
by a factor 3/2.
\bibitem{su3traceanomaly} M.Chanowitz and J.Ellis, Phys. Rev.
{\bf D7:2490},1973.
\bibitem{highordertrace} S.Adler, J.Collins, A.Duncan,  Phys.
Rev.{\bf D15,1712},1977; J.Collins, A.Duncan, S.Joglekar, Phys.
Rev.{\bf D16:438},1977; N.K.Nielsen, Nucl. Phys. {\bf B120:212},1977.
\bibitem{hm} K. Hagiwara and H. Murayama, Phys. Rev. {\bf D41:1001},1990.
\bibitem{rnc-mc} R.Cahn and M.Chanowitz, Phys.Rev.Lett. {\bf 56:1327},1986.
\bibitem{atlas} ATLAS Collaboration, CERN/LHCC/94-43 (Technical Proposal),
1994.
\bibitem{cms} CMS Collaboration, CERN/LHCC/94-38 (Technical Proposal), 1994.
\bibitem{qq-qqzz} U.Baur and E.Glover, Phys. Rev. {\bf D44:99},1991; V. Barger
et al., Phys. Rev. {\bf D44:1426},1991.
\bibitem{ohnemusetal} J. Ohnemus and J. Owens, Phys. Rev. {\bf D43:3626}, 1991.
\bibitem{wpwp} M. Berger and M. Chanowitz, Phys, Lett. {\bf B263:509}, 1991; V.
Barger et al., Phys. Rev. {\bf D42:3052}, 1990.
\bibitem{sdcloi} SDC Collaboration, E. Berger et al., SDC-90-00151, 1990
(unpublished).
\bibitem{mbmczz} M.Berger and M.Chanowitz, Phys. Rev. Lett. {\bf 68:757},1992.
\end{thebibliography}
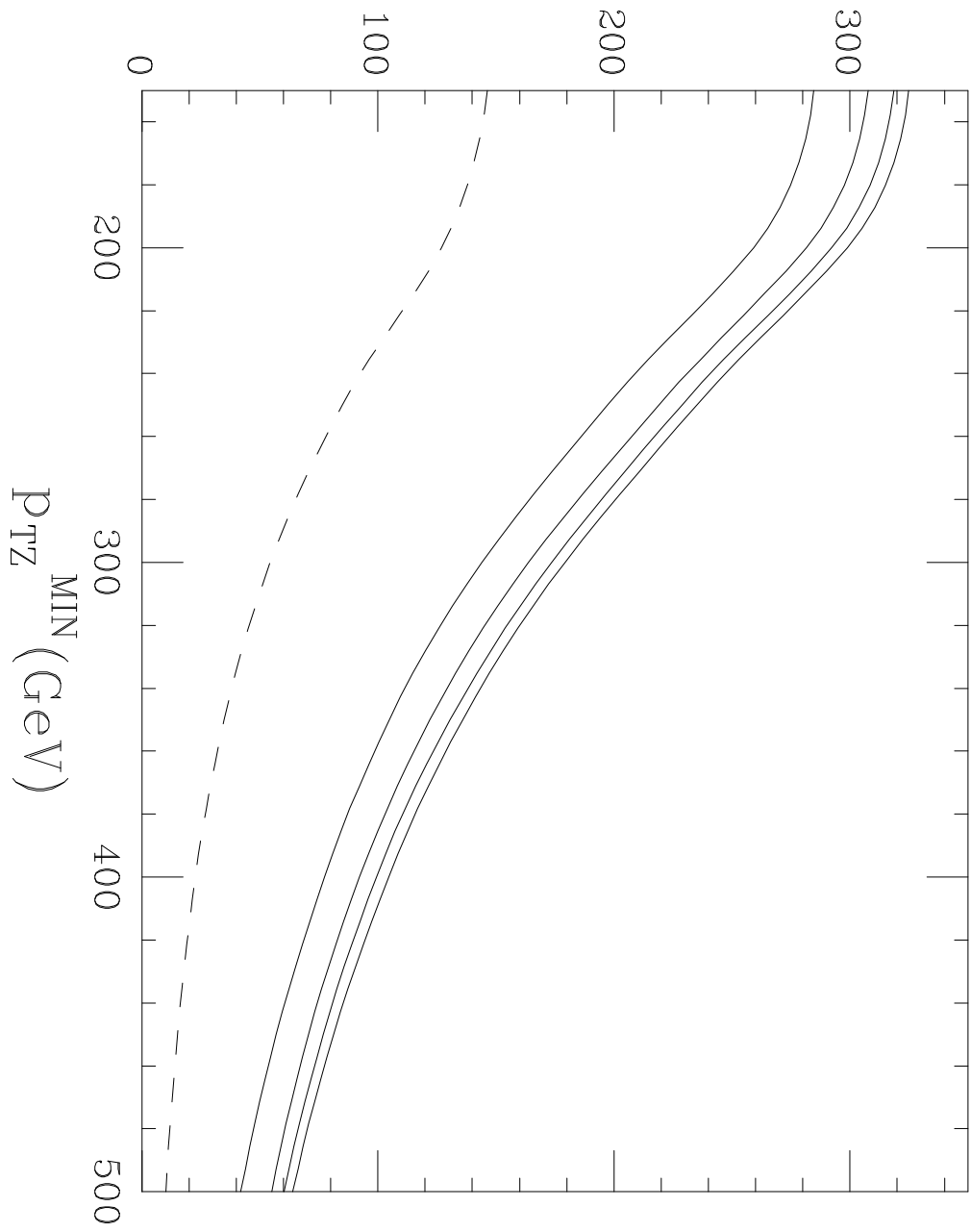
\end{document}